\pdfoutput=1
\documentclass[letterpaper, 10 pt, conference]{ieeeconf}  
\usepackage{amsmath,amsfonts}
\usepackage{algorithmic}
\usepackage{algorithm}
\usepackage{array}
\usepackage[caption=false,font=normalsize,labelfont=sf,textfont=sf]{subfig}
\usepackage{amsmath,amssymb}
\usepackage{textcomp}
\usepackage{stfloats}
\usepackage{url}
\usepackage{verbatim}
\usepackage{graphicx}
\usepackage{cite}
\usepackage{bm}
\usepackage{algorithm}
\usepackage{algorithmic}
\usepackage{empheq}
\usepackage{mathrsfs}
\usepackage{subcaption}
\usepackage{multirow}
\usepackage{booktabs}
\usepackage{subfig}

\newcommand{\bb}[1]{\mathbf{#1}}

\IEEEoverridecommandlockouts                              

\title{\LARGE \bf
Asynchronous Microphone Array Calibration using Hybrid TDOA Information
}

\author{Chengjie Zhang, Jiang Wang, and He Kong
\thanks{This work was supported by the Science, Technology, and Innovation Commission of Shenzhen Municipality, China, under Grant No. ZDSYS20220330161800001, the Shenzhen Science and
Technology Program under Grant No. KQTD20221101093557010, and the National Natural Science Foundation of China under Grant No. 62350055. The authors are with the Shenzhen Key Laboratory of Control Theory and Intelligent Systems, Southern University of Science and Technology, Shenzhen 518055, China. Emails: 12332644@mail.sustech.edu.cn; 12132297@mail.sustech.edu.cn; kongh@sustech.edu.cn
}}

\begin{document}

\maketitle
\thispagestyle{empty}
\pagestyle{empty}

\begin{abstract}
Asynchronous microphone array calibration is a prerequisite for many audition robot applications. A popular solution to the above calibration problem is the batch form of Simultaneous Localisation and Mapping (SLAM), using the time difference of arrival measurements between two microphones (TDOA-M), and the robot (which serves as a moving sound source during calibration) odometry information. In this paper, we introduce a new form of measurement for microphone array calibration, i.e. the time difference of arrival between adjacent sound events (TDOA-S) with respect to the microphone channels. We propose to use TDOA-S and TDOA-M, called hybrid TDOA, together with odometry measurements for bath SLAM-based calibration of asynchronous microphone arrays. Extensive simulation and real-world experiments show that our method is more independent of microphone number, less sensitive to initial values (when using off-the-shelf algorithms such as Gauss-Newton iterations), and has better calibration accuracy and robustness under various TDOA noises. Simulation results also demonstrate that our method has a lower Cramér-Rao lower bound (CRLB) for microphone parameters. To benefit the community, we open-source our code and data at https://github.com/AISLAB-sustech/Hybrid-TDOA-Calib.
\end{abstract}

\section{INTRODUCTION}
Microphone arrays can equip robots with sound source localization and tracking abilities, etc \cite{soundLocReview, Fu2024,strauss2018dregon,evers2020locata,lagace2023ego}. A prerequisite for realizing the above functionalities is to calibrate the array geometric information accurately \cite{GeoCalibReview}. A common approach to the above calibration problem is to utilize the time difference of arrival between microphone pairs (TDOA-M) from a series of sound events to estimate both microphone and sound source locations. Earlier methods require the clock synchronization of all microphones \cite{bilinearTOA, completeTOA}. To overcome the limitation, recent studies, including \cite{Ono, Wang, Badaway}, take into account the initial time offset between microphone channels.

\begin{figure}[htbp]
\centering
\includegraphics[scale=0.3]{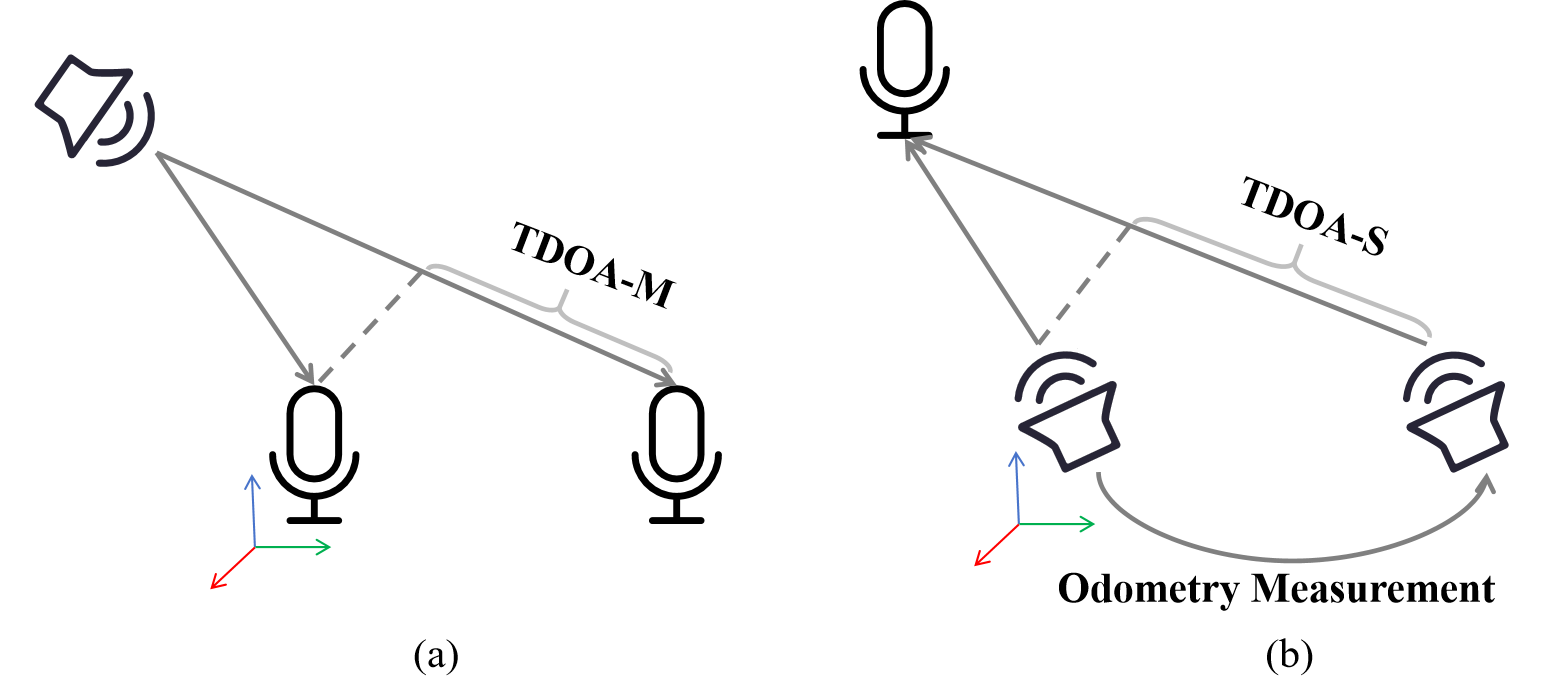}
\caption{Differences between TDOA-M (a) and TDOA-S (b). 
}
\label{fig:two-tdoas}
\end{figure}

During calibration, one can obtain the relative position measurements between adjacent sound events from the odometer onboard the robot (which acts as a moving sound source) and use them to improve the calibration accuracy. Following the above idea, based on TDOA-M and odometry measurements, an extended Kalman filter-based simultaneous localization and mapping (EKF-SLAM) method is proposed in \cite{Miura} to estimate microphone positions, time offsets, and sound source positions simultaneously. However, the impact of the other asynchronous factor, i.e. clock drift rate, is not considered. In \cite{SuKong, Su}, a batch SLAM-based method (see, e.g., \cite{graphSLAM}) is presented to jointly estimate microphone locations, time offsets, clock drift rates, and sound event positions. In our recent work, the framework in \cite{SuKong, Su} has also been generalized to multiple microphone arrays \cite{Wang2024} and refined by the data compression method \cite{Li2024}.

In this paper, we introduce a new form of measurement for microphone array calibration, i.e. the time difference of arrival between adjacent sound events (TDOA-S) with respect to the microphone channels. Fig. \ref{fig:two-tdoas} shows the difference between TDOA-M and TDOA-S in the calibration scene. We propose to combine TDOA-S with TDOA-M, called hybrid TDOA, with odometry measurements for microphone array calibration. Our main contributions are stated as follows.

\begin{itemize}
    \item [$\bullet$]
      We introduce TDOA-S and present a simple method to extract TDOA-S from raw audio data. To our best knowledge, this is the first time TDOA-S has been proposed in the literature and used in calibrating robot audition systems. The idea can be useful for other sensing modalities.
  \item [$\bullet$]
     We propose a batch SLAM-based calibration method utilizing hybrid TDOA information and odometry measurements to jointly estimate the asynchronous microphone array parameters (microphone positions, time offsets, clock drift rates) and sound source positions. Extensive simulations and real-world experiments show that compared to \cite{SuKong}, our method is more independent of microphone number, less sensitive to initialization, has higher accuracy and robustness under various TDOA noises, and has lower CRLB for microphone parameter estimates.
\end{itemize}


\section{THE PROPOSED METHOD}
Assume there are $N$ microphones. Denote the $i$-th microphone location, time offset, and clock drift rate as $\bb{x}_i$, $\tau_i$, and $\delta_i$ respectively. There are $K$ sound events and the $j$-th sound event location is  $\bb{s}_j$. Without loss of generality, the coordinate frame is established by sound event positions, called $Sound$ frame, $\bb{s}_1=\bb{0},(\bb{s}_2)_y=(\bb{s}_2)_z=(\bb{s}_3)_z=0$. The unknown microphone parameters in the $Sound$ frame are
\begin{equation}
\label{x_mic}
    \bb{x}_{mic}=[\bb{x}_1,\delta_1,\bb{x}_2,\tau_{2,1},\delta_2, ...,\bb{x}_N,\tau_{N,1},\delta_N]^T,
\end{equation}
where $\tau_{i,1}=\tau_i-\tau_1$, $i>1$. Sound source parameters that need to be estimated are
\begin{equation}
\label{s}
    \bb{s}=[(\bb{s}_2)_x,(\bb{s}_3)_x,(\bb{s}_3)_y,\bb{s}_4,...,\bb{s}_K]^T.
\end{equation}
\subsection{TDOA-S Derivation and Extraction}
\subsubsection{Derivation}
 TDOA-S is derived from the time of arrival (TOA) model that considers two asynchronous parameters: time offset and clock drift rate in microphones. In the absence of noise, the arrival time detected by $i$-th microphone for the $j$-th sound event, $T_{i,j}$ is shown below
\begin{equation}
\label{eq:toa}
    T_{i,j}=(1+\delta_i)(\frac{||\bb{x}_i-\bb{s}_j||}{c}+\tau_i+t_j),
\end{equation}
where $c$ is the known sound speed and $t_j$ is the emitting time of $j$-th sound event. $\tau_i$ and $\delta_i$ represent the shift and scaling of the temporal frame of $i$-th microphone with respect to (w.r.t.) the absolute temporal frame, respectively. The former is caused by different startup moments in different microphones. The latter is caused by the sampling rate mismatch between the microphone's actual and absolute sampling rates \cite{samplemismatch}, which can be modeled as a scale constant between a microphone's temporal frame and the absolute temporal frame. If the mismatch does not exist for $i$-th microphone ($\delta_i=0$), the TOA model is the same as the common TOA that only considers time offset \cite{GeoCalibReview}.


In indoor calibration scenarios, the distance between the microphone and sound events does not generally exceed 10 meters. In most cases (Table  \uppercase\expandafter{\romannumeral1} in \cite{SuKong}), the clock drift rate and time offset are less than $10^{-4}$ and 0.1s respectively. Therefore, the term $\delta_i(\frac{||\bb{x}_i-\bb{s}_j||}{c}+\tau_i)$ is negligible and can be ignored. After this simplification, $T_{i,j}$ becomes
\begin{equation}
\label{eq:toa2}
    \Tilde{T}_{i,j}=\frac{||\bb{x}_i-\bb{s}_j||}{c}+\tau_i+(1+\delta_i)t_j.
\end{equation}
Therefore, TDOA-S is expressed as $T^{S}_{i,j}=\Tilde{T}_{i,j+1}-\Tilde{T}_{i,j}$. The measurement model of $T^{S}_{i,j}$ is
\begin{equation}
\label{eq:tdoa-s}
T^{S}_{i,j}=\frac{||\bb{x}_i-\bb{s}_{j+1}||-||\bb{x}_i-\bb{s}_j||}{c}+(1+\delta_i)\Delta t_j,
\end{equation}
where $j<K$ and $\Delta t_j=t_{j+1}-t_j$ can be obtained accurately since the speaker installed on the robot is controllable.


\subsubsection{Extraction}
\begin{figure}[htbp]
\centering
\includegraphics[scale=0.3]{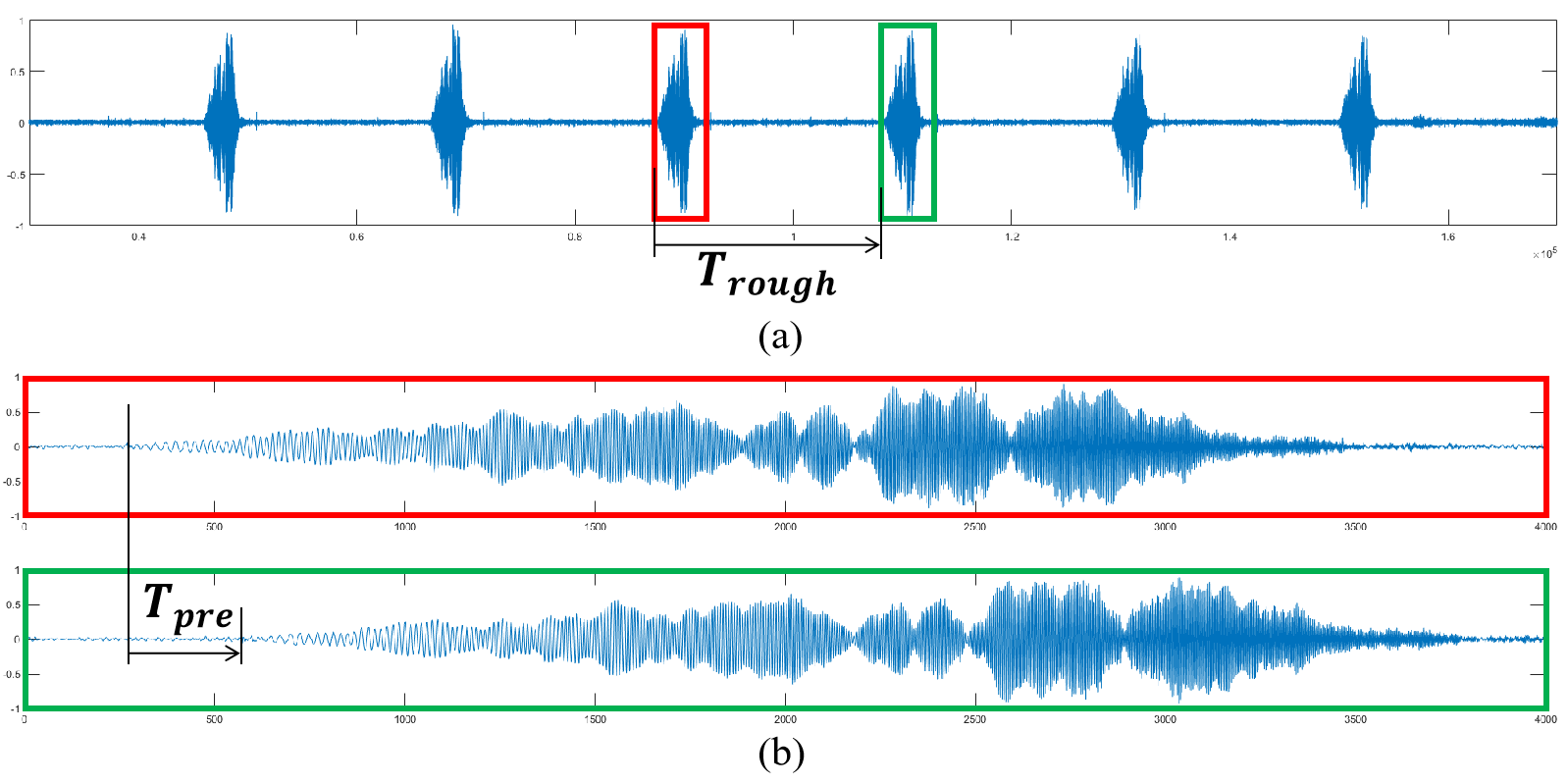}
\caption{Visualization of acquiring the rough delay: $T_{rough}$ (a) and precise delay: $T_{pre}$ (b). The red/green box represents the capture window obtaining the current/next recorded calibration signal.}
\label{fig:get-tdoa-s}
\end{figure}

There are two steps in obtaining TDOA-S, each visualized in Fig. \ref{fig:get-tdoa-s}. Initially, short-time energy \cite{introDSP} is used to obtain the rough left endpoint of each calibration signal in a single-channel microphone. The rough time delay of adjacent calibration signals ($T_{rough}$) is equal to the difference between the corresponding rough left endpoints (Fig. \ref{fig:get-tdoa-s}a). Next, each window containing a calibration signal is extracted based on the signal length and rough left endpoint, and we align the adjacent windows and perform GCC-PHAT \cite{GCC-PHAT} to obtain the precise delay ($T_{pre}$) (see Fig. \ref{fig:get-tdoa-s}b). Finally, one combines the rough delay and precise delay to obtain the overall delay ($T_{rough}+T_{pre}$), which is equal to the difference between two consecutive moments of arrivals.

\subsection{Calibration using Hybrid TDOA}
\subsubsection{Hybrid TDOA Measurements} The TDOA-S formulation is derived in (\ref{eq:tdoa-s}) and here we also derive the TDOA-M model based on (\ref{eq:toa2}). If we select the first microphone as a reference, TDOA-M becomes $T^{M}_{i,j}=\Tilde{T}_{i,j}-\Tilde{T}_{1,j}$,
\begin{equation}
\label{eq:tdoa-m}
    T^{M}_{i,j}=\frac{||\bb{x}_i-\bb{s}_j||-||\bb{x}_1-\bb{s}_j||}{c}+\tau_{i,1}+\delta_{i,1}t_j,
\end{equation}
where $\delta_{i,1}=\delta_i-\delta_1$ ($i>1$). Assume $t_1=0$ without loss of generality, for $j>1$, $t_j=t_j-t_1=\sum^{j}_{k=2}\Delta t_{k-1}$. The TDOA-M formula (\ref{eq:tdoa-m}) is equivalent to the TDOA formula in \cite{SuKong}. Hence, without noise, the total hybrid TDOA measurements are
\begin{equation}
    \bb{T}^H=[\bb{T}^S,\bb{T}^M]^T,
\end{equation}
where $\bb{T}^S=[\bb{T}^S_1,\bb{T}^S_2,...,\bb{T}^S_N]^T$, $\bb{T}^S_i=[T^S_{i,1},T^S_{i,2},...,T^S_{i,K-1}]^T$ and $\bb{T}^M=[\bb{T}^M_1,\bb{T}^M_2,...,\bb{T}^M_K]^T$, $\bb{T}^M_j=[T^M_{2,j},T^M_{3,j},...,T^M_{N,j}]^T$.

Considering i.i.d Gaussian noises, the real TDOA-M and TDOA-S measurements are $t^{M}_{i,j}=T^{M}_{i,j}+w^{M}_{i,j}$ ($i>1$) and  $t^{S}_{i,j}=T^{S}_{i,j}+w^{S}_{i,j}$ ($j<K$), respectively, with $w^{M}_{i,j},w^{S}_{i,j}\sim N(0,\sigma^2_{tdoa})$. The real hybrid TDOA measurements are
\begin{equation}
    \bb{t}^H=[\bb{t}^S,\bb{t}^M]^T,
\end{equation}
where $\bb{t}^S=[\bb{t}^S_1,\bb{t}^S_2,...,\bb{t}^S_N]^T$, $\bb{t}^S_i=[t^S_{i,1},t^S_{i,2},...,t^S_{i,K-1}]^T$ and $\bb{t}^M=[\bb{t}^M_1,\bb{t}^M_2,...,\bb{t}^M_K]^T$, $\bb{t}^M_j=[t^M_{2,j},t^M_{3,j},...,t^M_{N,j}]^T$. 

Under Gaussian noise $\bb{v}_j\sim N(\bb{0},\sigma^2_{odo}\bb{I_3})$, the odometry measurements are $\bb{m}=[\bb{m}_1,\bb{m}_2,..., \bb{m}_{K-1}]^T$ with $\bb{m}_j$ being defined as follows
\begin{equation}
\label{eq:odo}
    \bb{m}_j=\Delta\bb{s}_j+\bb{v}_j=\bb{s}_{j+1}-\bb{s}_j+\bb{v}_j,
\end{equation}
where $j<K$. Denote $\Delta\bb{s}=[\Delta\bb{s}_1,\Delta\bb{s}_2, ..., \Delta\bb{s}_{K-1}]^T$.
\subsubsection{Nonlinear Least Squares solved by Gauss-Newton}
From the perspective of batch SLAM, nodes are the locations of a series of sound events (robot pose without orientation) and microphone array (landmark) with positions and asynchronous parameters, while edges are odometry measurements and hybrid TDOA measurements. Note that during the calibration process, any microphone observes every sound event.
One can then construct the corresponding nonlinear least squares based on maximum likelihood estimate (MLE) and then use the Gauss-Newton (GN) method to estimate microphone array positions, time offsets, clock drift rates, and the sound event locations simultaneously. Denote the unknown parameters $\bb{x}=[\bb{x}_{mic},\bb{s}]^T$, measurements 
$\bb{z}=[\bb{t}^H,\bb{m}]^T$ and measurement function $\bb{f}(\bb{x})=[\bb{T}^H,\Delta \bb{s}]^T$. The minimum of the nonlinear least squares (LS) is shown below\\
\begin{equation}
    \label{LS}
    \min_{\bb{x}} \ (\bb{f}(\bb{x})-\bb{z})^T\bb{W}^{-1}(\bb{f}(\bb{x})-\bb{z}),
\end{equation}
where $\bb{W}=diag(\sigma^2_{tdoa}\bb{I}_{N(K-1)+K(N-1)},\sigma^2_{odo}\bb{I}_{3K-3})$. For solving the above optimization problem, the GN method is usually used. Moreover, for performing source localization tasks after calibration, we need to convert $\bb{x}_{mic}$ in $Sound$ frame to $\bb{x}_{mic}$ in $Mic.$ frame. The details of the transformation are shown in Appendix A.

\subsection{Computation of CRLB}
CRLB is a popular and powerful tool for analyzing parameter estimation errors, as it provides a lower bound on the estimated parameter variance for any unbiased estimator. 
For nonrandom vector parameters, the CRLB states that the covariance matrix of an unbiased estimator is bounded as follows \cite{estimation},
\begin{equation}
    E[(\hat{\bb{x}}(\bb{z})-\bb{x}_0)(\hat{\bb{x}}(\bb{z})-\bb{x}_0)^T]\geq \bb{C},
\end{equation}
where $\hat{\bb{x}}(\bb{z})$ is an unbiased estimator of $\bb{x}$ given measurement $\bb{z}$, $\bb{x}_0$ is the true value of vector parameter of $\bb{x}$ and $\bb{C}$ is the CRLB matrix w.r.t. parameters $\bb{x}$. $\bb{C}=\bb{F}^{-1}$ and $\bb{F}$ is the Fisher information matrix,
\begin{equation}
    \bb{F}=E[[\triangledown_{\bb{x}} ln \bb{L}(\bb{x})][\triangledown_{\bb{x}} ln \bb{L}(\bb{x})]^T]|_{\bb{x}=\bb{x}_0}.
\end{equation}
Furthermore, the Fisher information matrix is shown below,
\begin{equation}
    \bb{F}=\bb{J}^T\bb{W}^{-1}\bb{J}.
\end{equation}
In our method, we consider the GN solver for the nonlinear least squares (\ref{LS}) as the unbiased estimator and the CRLB matrix of $\bb{x}_{mic}$ in $Sound$ frame, called $\bb{x}^S_{mic}$ here, is defined as $\bb{C}_{\bb{x}_{mic}^S}=\bb{C}(1:5N-1,1:5N-1)$, which is the submatrix of $\bb{C}$ w.r.t. $\bb{x}$. Then, we need to obtain CRLB for $\bb{x}_{mic}$ in $Mic.$ frame, called $\bb{x}^M_{mic}$. The affine transformation between $\bb{x}^M_{mic}$ and $\bb{x}^S_{mic}$ is represented below
\begin{equation}
    \bb{x}_{mic}^M=\bb{A}^M_S \bb{x}_{mic}^S + \bb{b}^M_S,
\end{equation}
where the expression of $\bb{A}^M_S$ and $\bb{b}^M_S$ are shown in Appendix A.
According to \cite{estimationTheory} in Section 3.8, the CRLB matrix of $\bb{x}_{mic}^M$, $\bb{C}_{\bb{x}_{mic}^M}$ is shown below,
\begin{equation}
    \bb{C}_{\bb{x}_{mic}^M}=\bb{A}^M_S \bb{C}_{\bb{x}_{mic}^S} (\bb{A}^M_S)^T.
\end{equation}
In $\bb{C}_{\bb{x}_{mic}^M}$, we extract diagonal elements corresponding to the CRLB for $\bb{x}_{mic}^M$. Then we define an indicator $D_{CRLB}$
\begin{equation}
    D_{CRLB}=\sqrt{\frac{\sum_{i=2}^{N} \ D_{CRLB_i}}{N-1}},
\end{equation}
where $D_{CRLB_i}$ is represented as CRLB for $i$-th microphone location, offset, or clock drift rate in $Mic.$ frame.

\section{SIMULATIONS}
We next present simulations to validate the advantages of our method: independence of the number of microphones (Part A), less insensitivity to initial values (Part B), better calibration accuracy and robustness under various TDOA noises (Part C), and lower CRLB for microphone parameters (Part D). For comparative analysis, we use the calibration method \cite{SuKong} using TDOA-M in 3D.

\subsubsection{Setup} We design three motion trajectories of a sound source. The first one has the space of 3m$\times$3m$\times$3m with 8 sound events (trajectory 1), the second has the space of 2m$\times$6m$\times$2m with 10 sound events (trajectory 2), and the third one possesses the space of 4m$\times$4m$\times$2m with 14 sound events (trajectory 3).
\begin{table}[htbp]
    \caption{\small SIMULATION SETTINGS} 
\begin{minipage}[t]{0.5\textwidth}
\centering
\begin{tabular}{|c|c|c|c|}
  \hline
  Setup & Part A & Part B & Part C/D \\
  \hline
  $N$ & 4,6,8,10 & 6 & 6\\
  \hline
  $K$ & 8/10/14 &  8/10/14 &  8/10/14\\
  \hline
  True $\bb{x}_{mic}$ & \multicolumn{3}{|c|}{random}\\
  \hline
  Initial $\bb{x}$ & random & $\sigma_{init}$ & random\\
  \hline
  $\sigma_{tdoa}$ & 0.1ms & 0.1ms & 0.05,0.1,0.5ms\\
  \hline
  $\sigma_{odo}$ & \multicolumn{3}{|c|}{0.01m}\\
  \hline
\end{tabular}
\label{sim_setup}
\end{minipage}
\end{table}

In ``True $\bb{x}_{mic}$'', ``random'' means microphone locations are randomly generated in the corresponding trajectory space and $|\tau_{i,1}|\leq0.1s$, $|\delta_i|\leq10^{-4}s$. In ``Initial $\bb{x}$'', ``random'' means both microphone and sound event locations are randomly generated in the corresponding trajectory space, and asynchronous parameters are set to be zero. $\sigma_{init}$ are standard deviations (SDs) of zero-mean Gaussian noises added into the true positions as the initial values of both microphone and sound event locations. In trajectory 1, $\sigma_{init}=0m, 1m, 2m, 3m$ and in trajectory both 2 and 3, $\sigma_{init}=0m, 2m, 4m, 6m$. Simulation under different numbers of microphones (Part A), various initial value noises (Part B), and several TDOA noises (Part C/D) repeat 200 times in each trajectory and the results of three trajectories are combined to analyze.

\subsubsection{Evaluation Metric} The average root mean square errors of the estimated microphone locations (Loc. err.), time offset (Off. err.), and clock drift rates (Dri. err.) are evaluated in the $Mic.$ frame, whose definition is in Appendix A.


\begin{figure*}[htbp]
\centering
\includegraphics[scale=0.25]{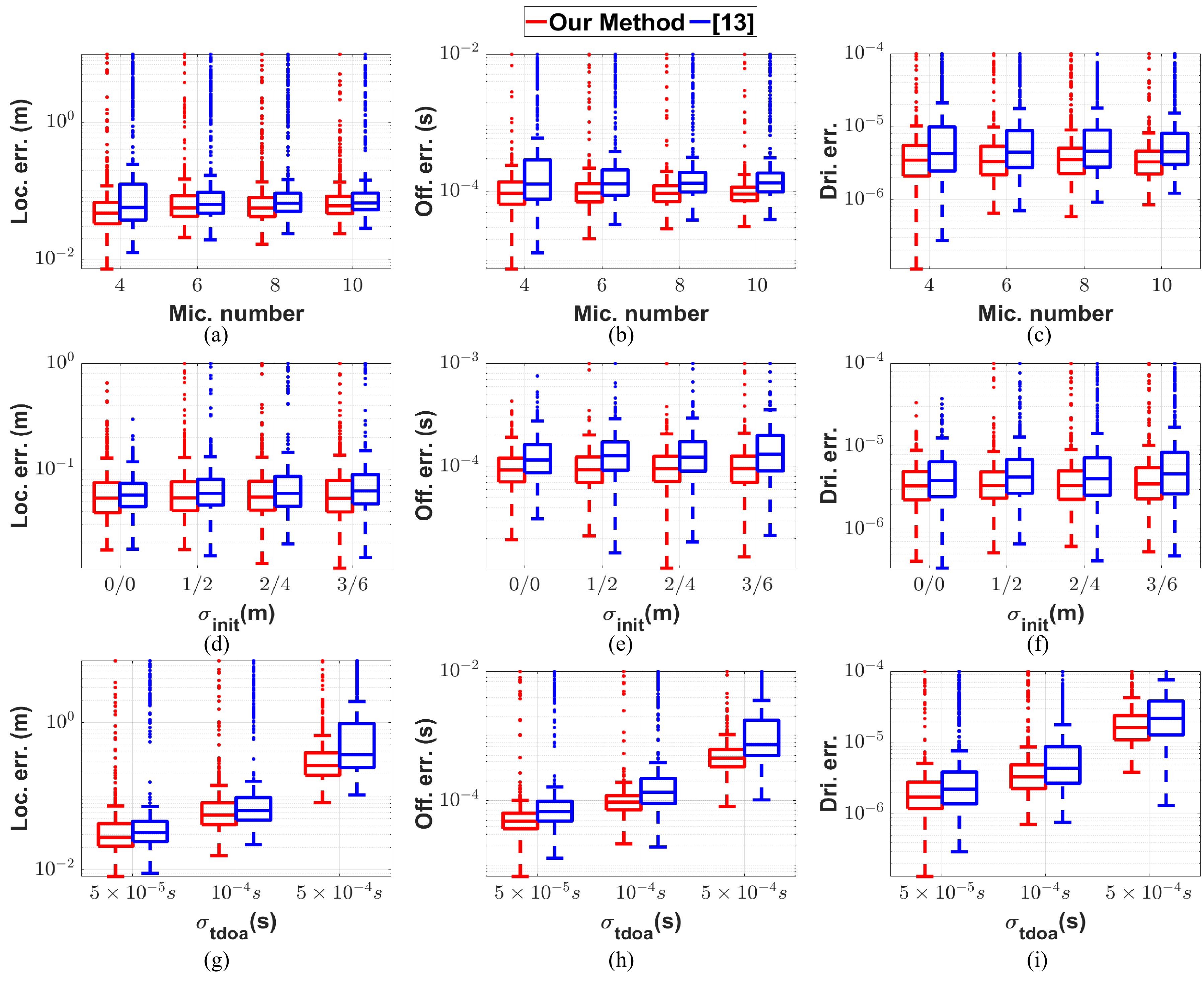}
\caption{Box plot of estimation errors for microphone parameters in simulations: microphone locations (a),(d),(g), time offsets (b),(e),(h), and clock drift rates (c),(f),(i) under four microphone numbers, i.e. 4, 6, 8, and 10, four initial values noise SDs under three trajectories, i.e. 0m/0m, 1m/2m, 2m/4m, and 3m/6m, and three TDOA noise SDs, i.e. $5\times10^{-5}s$, $10^{-4}s$, and $5\times10^{-4}s$, respectively. Here, ``$x$m/$y$m'' means combining the estimation result of trajectory 1 under $\sigma_{init}=x$m and results of both trajectory 2 and 3 under $\sigma_{init}=y$m.}
\label{fig:all_sim}
\end{figure*}

\subsubsection{Results}
It can be observed from Fig. \ref{fig:all_sim}a-c that as the number of microphones changes, the calibration performance for microphone parameters of our method remains basically unchanged. However, the performance of \cite{SuKong} shows significant changes and approaches that of our method as the number increases. Fig. \ref{fig:all_sim}d-f shows that the estimation performance for microphone parameters of our method remains unchanged under different initial value noises. However, microphone parameters estimated by \cite{SuKong} exhibit an increase in estimation error as the initial values noise increases. In Fig. \ref{fig:all_sim}g-i, we can observe that our method has better accuracy and robustness in estimating microphone parameters under three levels of TDOA noises, as we have lower median and interquartile range (IQR) values for each box. Table \uppercase\expandafter{\romannumeral2} confirms that our method estimates the CRLB for microphone parameters to be smaller under various TDOA noises.

\begin{table}[htbp]
    \caption{\small CRLB results under various TDOA noises (Bold means better)} 
    \begin{minipage}[t]{0.3\textwidth}
    \centering
    \resizebox{1.62\textwidth}{!}{
    \begin{tabular}{cccc}
      \toprule
      $\sigma_{tdoa}=5\times10^{-5}s$ & Loc. err. (m) & Off. err. (ms) & Dri. err. ($10^{-6}$) \\
      \midrule
      \cite{SuKong} & 0.033 & 0.074 & 2.765\\
      Our method & \textbf{0.027} &  \textbf{0.055} &  \textbf{2.191}\\
      \bottomrule
    \end{tabular}
    }
    \label{tab1}
    \end{minipage}
    \vfill
    \begin{minipage}[t]{0.3\textwidth}
    \centering
    \resizebox{1.62\textwidth}{!}{
    \begin{tabular}{cccc}
      \toprule
      $\sigma_{tdoa}=1\times10^{-4}s$ & Loc. err. (m) & Off. err. (ms) & Dri. err. ($10^{-6}$)\\
      \midrule
      \cite{SuKong} & 0.064 & 0.140 & 5.148\\
      Our method &  \textbf{0.044} &  \textbf{0.102} &  \textbf{3.989}\\
      \bottomrule
    \end{tabular}
    }
    \label{tab2}
    \end{minipage}
    \vfill
    \begin{minipage}[t]{0.3\textwidth}
    \centering
    \resizebox{1.62\textwidth}{!}{
    \begin{tabular}{cccc}
      \toprule
      $\sigma_{tdoa}=5\times10^{-4}s$ & Loc. err. (m) & Off. err. (ms) & Dri. err. ($10^{-6}$)\\
      \midrule
      \cite{SuKong} & 0.310 & 0.673 & 24.169\\
      Our method &  \textbf{0.191} &  \textbf{0.488} &  \textbf{18.681}\\
      \bottomrule
    \end{tabular}
    }
    \label{tab2}
    \end{minipage}
\end{table}


\section{REAL-WORLD EXPERIMENT}

\subsubsection{Calibration Scenario} 
\begin{figure}[htbp]
\centering
\includegraphics[scale=0.2]{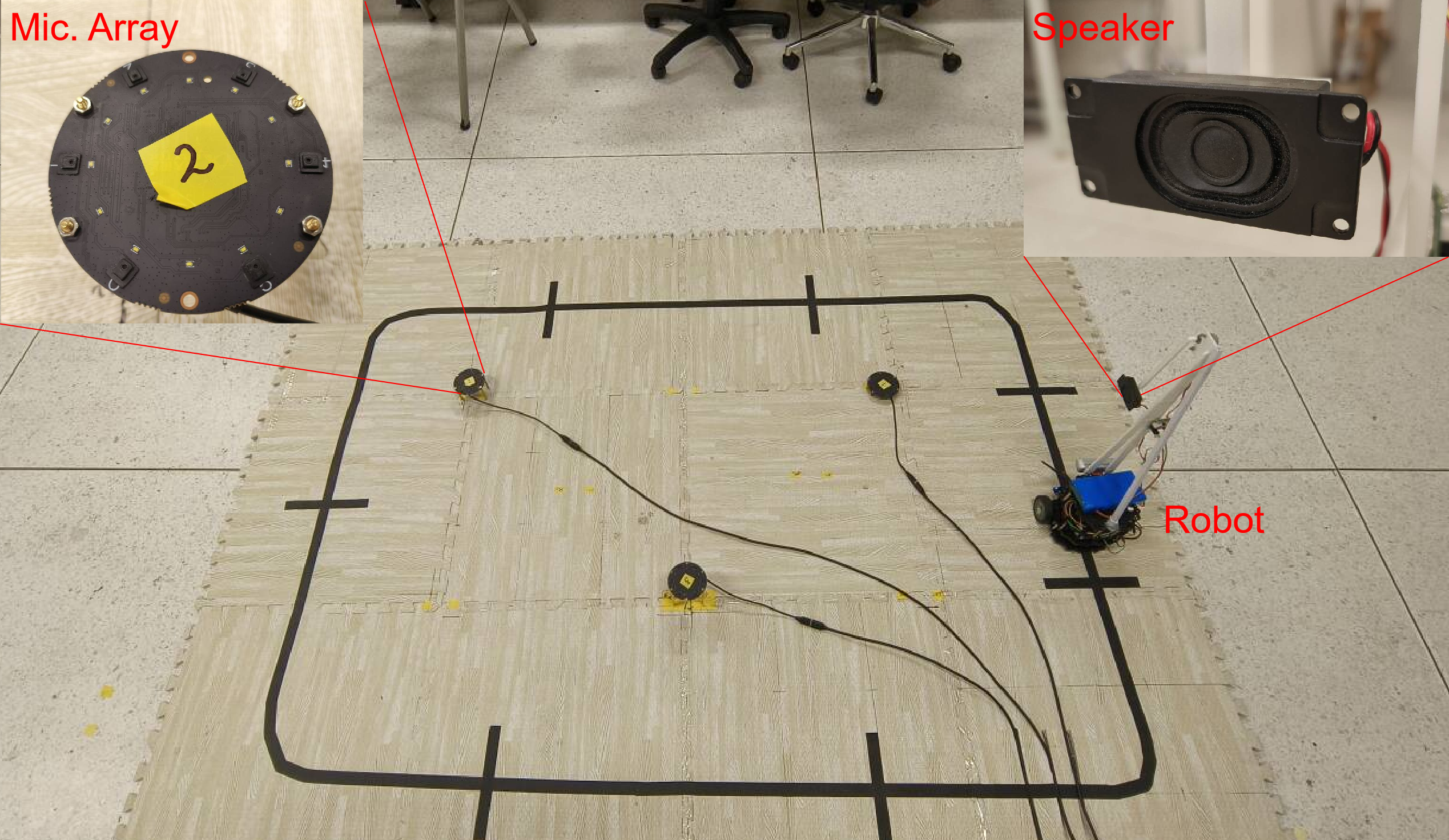}
\caption{The calibration scenario for real-world experiments.}
\label{fig:realscene}
\end{figure}

The real-world calibration scenario is shown in Fig. \ref{fig:realscene}. The robot (TurtleBot3) carrying a speaker moves around a given plane trajectory whose space is $1.6m\times2m\times1m$. When the robot reaches the marked point, the speaker sends out a calibration signal (chirp), and there are 14 sound event locations. On the robot, the speaker is installed on a rotatable pole to change the height of the sound source. Both TDOA-S and TDOA-M are obtained by the GCC-PHAT method \cite{GCC-PHAT} and odometry measurements are obtained by an efficient Monocular Visual-Inertial State Estimator (VINS-Mono) \cite{vins_mono}. There are three microphone arrays inside the trajectory, each array uses IFLYTEK M160C, a circular array with six microphones.

\subsubsection{Setup} We randomly set five microphone position configurations and each one is repeated three times. A certain number of microphones are selected from the three arrays randomly to form a microphone array. The advantage of extracting microphones from multiple arrays to form an array is that it can generate a large amount of real data more conveniently. We conduct similar comparisons with Section III (Part A, Part B, and Part C). The real-world experiment settings are the same as shown in Table \ref{sim_setup}, except that there is only one sound source trajectory with 14 sound events, and TDOA noises of real-world data need to be derived based on the real data. Also, in ``True $\bb{x}_{mic}$'', ``random'' means microphones are randomly selected from three microphone arrays. $\sigma_{tdoa}$ is set to $10^{-4}s$.

\subsubsection{TDOA Noises Evaluation}
It's necessary to estimate the noises of TDOA-S and TDOA-M before conducting the real experiment. Because the true values of both microphone and sound locations are known, the estimated noise standard deviation of TDOA-S ($\Tilde{\sigma}^S_{tdoa}$) and TDOA-M ($\Tilde{\sigma}^M_{tdoa}$) are obtained based on MLE in Appendix B. In Part A and B, to ensure fairness, we select data satisfying $|\Tilde{\sigma}^S_{tdoa}-\Tilde{\sigma}^M_{tdoa}|<10^{-5}s$. In Part C, data is divided into five cases with different estimated TDOA noises: $\Tilde{\sigma}^S_{tdoa},\Tilde{\sigma}^M_{tdoa}<10^{-4}s$ (Case A), $10^{-4}s<\Tilde{\sigma}^S_{tdoa}, \Tilde{\sigma}^M_{tdoa}<1.5\times10^{-4}s$ (Case B), $1.5\times10^{-4}s<\Tilde{\sigma}^S_{tdoa}, \Tilde{\sigma}^M_{tdoa}<5\times10^{-4}s$ (Case C), $|\Tilde{\sigma}^S_{tdoa}-\Tilde{\sigma}^M_{tdoa}|<10^{-5}s$ (Case D) and all TDOA-S and TDOA-M without any conditions (Case E).

\subsubsection{Results} 
\begin{figure*}[htbp]
\centering
\includegraphics[scale=0.29]{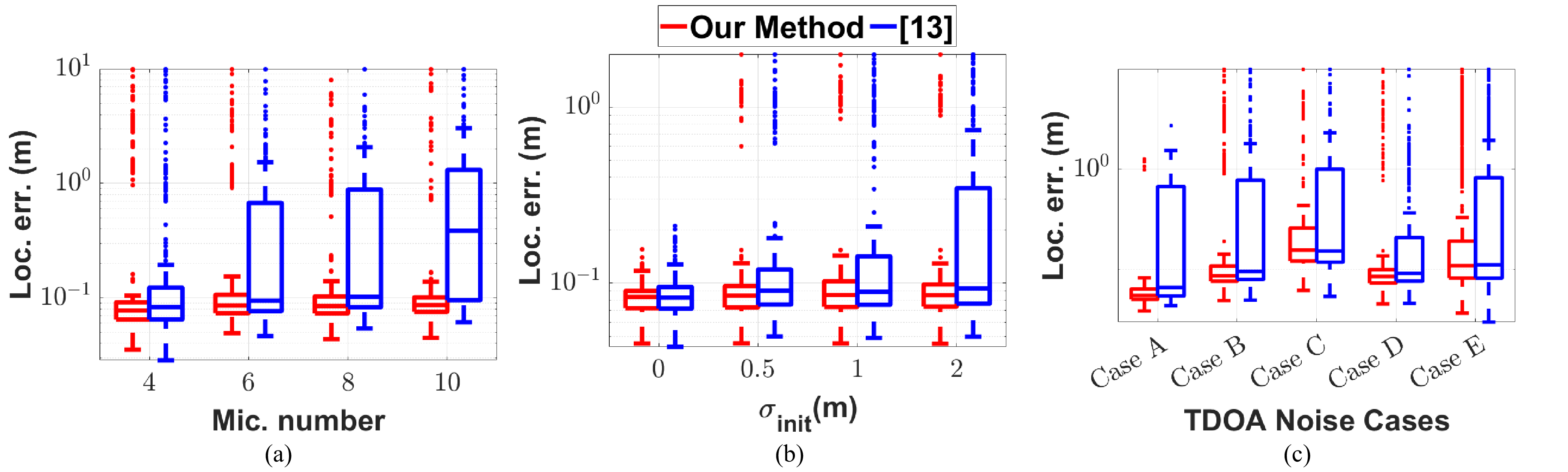}
\caption{Box plot of results in real-world experiment: microphone location estimation errors under various microphone numbers (a), i.e. 4, 6, 8, and 10, initial values noise SDs (b), i.e. 0m, 0.5m, 1m, and 2m, and five cases of TDOA noises (c).}
\label{fig:real1-3}
\end{figure*}

Fig. \ref{fig:real1-3} shows microphone location estimation results in real-world experiments and proves our method performs well and independently of the number of microphones (Fig. \ref{fig:real1-3}a), has low sensitivity to initial values (Fig. \ref{fig:real1-3}b), and is robust under different TDOA noise levels (Fig. \ref{fig:real1-3}c). In Case C of Fig. \ref{fig:real1-3}c, although the accuracy of our method is slightly lower than \cite{SuKong} due to the average $\Tilde{\sigma}^S_{tdoa}$ is 80$\mu$s larger than that of $\Tilde{\sigma}^M_{tdoa}$, our method remains much more stable with a smaller IQR.

\section{DISCUSSION}
The simulation and experimental results show that our calibration method, incorporating both TDOA-S and TDOA-M measurements, has several advantages compared to the case using only TDOA-M information. A major reason for the above outcome could be that TDOA-S actually has fewer timing parameters than TDOA-M, which helps to reduce the nonlinearity of non-linear LS and sensitivity to initial values. For lower CRLB, from an information theory perspective, compared to \cite{SuKong}, we have more measurement information, which results in lower CRLB.



\section{CONCLUSIONS}
This paper is concerned with asynchronous microphone array calibration using batch SLAM. More specifically, we have introduced a new type of measurement, i.e. TDOA-S, for the above calibration problem. We have presented a simple procedure to extract TDOA-S measurements from raw audio information. We proposed to use hybrid TDOA information (both TDOA-S and TDOA-M) for calibrating asynchronous microphone arrays. 
Both simulation and experiment results show that our proposed method leads to improved calibration results compared to the case using only TDOA-M information. In particular, the proposed method is more independent of the number of microphones, has lower sensitivity to initial values, and has higher accuracy and robustness under various TDOA noises. The focus of our current and future work is to explore the relationship between calibration observability and sound source trajectory configuration and generalize the ideas in this paper to the problem of calibrating multiple microphone arrays.

\section{APPENDIX}
\subsection{Affine Transformation from $\bb{x}_{mic}^S$ to $\bb{x}_{mic}^M$}
$\bb{x}_{mic}$ in $Mic.$ frame which is established by assuming $\bb{x}_1=\bb{0}$, $(\bb{x}_2)_y=(\bb{x}_2)_z=(\bb{x}_3)_z=0$, is defined below:
\begin{equation}
    \label{x_mic2}
    \bb{x}_{mic}=[\bb{x}_1,\bb{x}_2,\tau_{2,1},\delta_{2,1}, ...,\bb{x}_N,\tau_{N,1},\delta_{N,1}]^T.
\end{equation}
Given the definitions of the vectors $\bb{x}_{mic}^S$ and $\bb{x}_{mic}^M$, the details of this linear transformation relationship are as follows
\begin{equation}
    \label{gt_transform}
    \begin{split}
    \bb{x}^M_i&=\bb{R} \bb{x}^S_i + \bb{t},\\
    \tau^M_{i,1}&=\tau^S_{i,1},\\
    \delta^M_{i,1}&=\delta^S_i-\delta^S_1.\\
    \end{split}
\end{equation}
where $\bb{R}$ and $\bb{t}$ are the rotation matrix and translation vector respectively and transfer $\bb{x}_i$ from $Sound$ frame into $Mic.$ frame. The construction of $\bb{A}^M_S$ and $\bb{b}^M_S$ are based on (\ref{x_mic}), (\ref{x_mic2}) and (\ref{gt_transform}).

\subsection{Estimating Standard Deviation of TDOA Noise}
\subsubsection{Computation of $\Tilde{\sigma}^{S}_{tdoa}$} Given $t^{S}_{i,j}$, $\bb{x}_i$ and $\bb{s}_j$, $i=1,2,...,N$ and $j=1,2,...,K-1$. $\Tilde{t}^{S}_{i,j}$ is shown below
\begin{align*}
    \Tilde{t}^{S}_{i,j}&=t^{S}_{i,j}-\frac{||\bb{x}_i-\bb{s}_{j+1}||-||\bb{x}_i-\bb{s}_j||}{c}-\Delta t_j\\
    &=\delta_i\Delta t_j+w^S_{i,j}.
\end{align*}
Unbiased estimation based on MLE for $\delta_i$ is below
\begin{align*}
\min_{\delta_i} \ \sum_{j=1}^{K-1} (\Tilde{t}^S_{i,j}-\delta_i\Delta t_j)^2 
 \Longrightarrow \hat{\delta}_i=\frac{\sum_{j=1}^{K-1} \Tilde{t}^S_{i,j}}{\sum_{j=1}^{K-1} \Delta t_j}.
\end{align*}
Therefore, $\Tilde{w}^S_{i,j}=\Tilde{t}^S_{i,j}-\hat{\delta}_i\Delta t_j$. $\Tilde{\sigma}^{S}_{tdoa}$ is estimated unbiased based on $\Tilde{w}^S_{i,j}$.

\subsubsection{Computation of $\Tilde{\sigma}^{M}_{tdoa}$} Given $t^{M}_{i,j}$, $\bb{x}_i$ and $\bb{s}_j$, $i=2,3,...,N$ and $j=1,2,...,K$. $\Tilde{t}^{M}_{i,j}$ is shown below
\begin{align*}
    \Tilde{t}^{M}_{i,j}=t^{M}_{i,j}-\frac{||\bb{x}_i-\bb{s}_j||-||\bb{x}_1-\bb{s}_j||}{c}=\tau_{i,1}+\delta_{i,1}t_j+w^M_{i,j}.
\end{align*}
Unbiased estimation based on MLE for $\tau_{i,1},\delta_{i,1}$ are below
\begin{align*}
\min_{\tau_{i,1},\delta_{i,1}} \ \sum_{j=1}^K (\Tilde{t}^{M}_{i,j}-\tau_{i,1}-\delta_{i,1}t_j)^2 
 \Longrightarrow \begin{bmatrix}
    \hat{\tau}_{i,1}\\
    \hat{\delta}_{i,1}
\end{bmatrix}=(A^T A)^{-1}A^T b,
\end{align*}
$A=\begin{bmatrix}
    1 & t_1\\
    1 & t_2\\
    \vdots & \vdots\\
    1 & t_K
\end{bmatrix}$ and $b=\begin{bmatrix}
    \Tilde{t}^{M}_{i,1}\\
    \Tilde{t}^{M}_{i,2}\\
    \vdots\\
    \Tilde{t}^{M}_{i,K}
\end{bmatrix}$. Therefore, $\Tilde{w}^M_{i,j}=\Tilde{t}^{M}_{i,j}-\hat{\tau}_{i,1}-\hat{\delta}_{i,1}t_j$. $\Tilde{\sigma}^{M}_{tdoa}$ is estimated unbiased based on $\Tilde{w}^M_{i,j}$.

\bibliographystyle{IEEEtran}
\bibliography{ref}

\end{document}